\documentclass{article}
\usepackage{siunitx}
\usepackage{graphicx}
\usepackage{authblk}
\usepackage{xcolor}
\usepackage{amsmath}
\usepackage{hyperref}

\def\draftmode{1}

\if\draftmode1

\newcommand{\Nate}[1]{\textbf{\textcolor{brown}{(Nate) #1}}}
\newcommand{\Derek}[1]{\textbf{\textcolor{teal}{(Derek) #1}}}
\newcommand{\Dima}[1]{\textbf{\textcolor{red}{(Dima) #1}}}
\newcommand{\Matthew}[1]{\textbf{\textcolor{blue}{(Matthew) #1}}}
\newcommand{\Note}[1]{\textcolor{violet}{#1}}

\else

\newcommand{\Nate}[1]{}
\newcommand{\Derek}[1]{}
\newcommand{\Dima}[1]{}
\newcommand{\Matthew}[1]{}
\newcommand{\Note}[1]{}

\fi

\begin{document}

\title{A network of superconducting gravimeters as a detector of matter with feeble nongravitational coupling}

\author[1]{Wenxiang Hu}
\affil[1]{School of Physics, Peking University, Beijing, China}

\author[2,3]{Matthew M. Lawson}
\affil[2]{The Oskar Klein Centre for Cosmoparticle Physics, Department of Physics, Stockholm University, AlbaNova, 10691 Stockolm, Sweden}
\affil[3]{Nordita, KTH Royal Institute of Technology and Stockholm University, Roslagstullbacken 23, 10691 Stockholm, Sweden}

\author[4,5]{Dmitry Budker}
\affil[4]{Helmholtz Institut, Johannes Gutenberg-Universit{\"a}t Mainz, 55128 Mainz, Germany}
\affil[5]{Department of Physics, University of California, Berkeley, California 94720, USA}

\author[4]{Nataniel~L.~Figueroa}

\author[6]{Derek~F.~Jackson~Kimball}
\affil[6]{Department of Physics, California State University - East Bay, Hayward, California, USA}

\author[7]{Allen~P.~Mills~Jr.}
\affil[7]{Department of Physics and Astronomy, University of California, Riverside, USA}

\author[8]{Christian~Voigt}
\affil[8]{GFZ German Research Center for Geosciences, Telegrafenberg, Potsdam, Germany}


\date{\today}

\setcounter{Maxaffil}{0}
\renewcommand\Affilfont{\itshape\small}

\maketitle

\begin{abstract}
\noindent Hidden matter that interacts only gravitationally would oscillate at characteristic frequencies when trapped inside of Earth. For small oscillations near the center of the Earth, these frequencies are around 300\,$\mu$Hz. Additionally, signatures at higher harmonics would appear because of the non-uniformity of Earth's density. In this work, we use data from a global network of gravimeters of the International Geodynamics and Earth Tide Service (IGETS) to look for these hypothetical trapped objects. We find no evidence for such objects with masses on the order of 10$^{14}$\,kg or greater with an oscillation amplitude of 0.1 $r_e$. It may be possible to improve the sensitivity of the search by several orders of magnitude via better understanding of the terrestrial noise sources and more advanced data analysis.
\end{abstract}

\section{Introduction}

A classic result in Newtonian gravity is that if a small mass is orbiting inside a
large mass of uniform density, such that the orbit is entirely contained in the
interior of the large mass, the period of the orbit is fixed by the density
of the large mass, and independent of the particulars of the orbit. This is because the system can be described as a three-dimensional harmonic oscillator. In the case of a mass inside a sphere with uniform density equal to the average density of the Earth, the period of such orbits
would be approximately 80 minutes. Such a scenario is impossible with masses comprised of ordinary matter because of nongravitational interactions. However, the situation could be, in fact, be hypothetically realized if the small mass is comprised of some ``hidden matter" (we call it a  hidden internal object, HIO) that has only feeble, if any, non-gravitational interactions with normal matter. Furthermore, it is known that such hidden matter exists: evidence from many independent observations point to the existence of dark matter \cite{Bertone2018}, an invisible substance which may interact with ordinary matter primarily via gravity. If some fraction of this hidden matter is gravitationally bound
within the Earth, this hypothetical scenario of a hidden internal object could be realized.

\begin{figure} 
\label{Fig:Earth Density Profile}
\centering
\includegraphics[width=\textwidth]{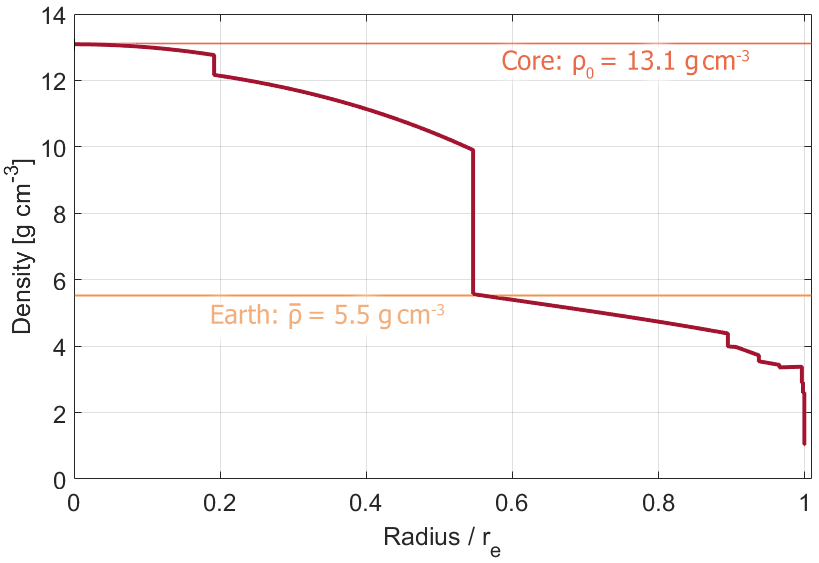}
\caption{The density profile of the Earth based on the Preliminary Reference Earth model (PREM) \cite{Dziewonski1981}. $r_{e}$\,=\,6371\,km is the mean radius of the Earth, $\bar{\rho}\,=$\,5.51\,g\,cm$^{\text{-3}}$ is the average density of the Earth, and $\rho_{0}=$\,13.1\,g\,cm$^{\text{-3}}$ is the density at the Earth's center.}
\end{figure}

This suggests a tantalizing scenario. Perhaps, one can detect the presence of such HIO via sensitive measurements of gravitational acceleration at the surface of the Earth. An attractive feature of this idea is that the method does not depend on any specifics of what the orbiting matter is composed of and, in the case of uniform density, would lead to a signal at a well defined frequency. Unfortunately, the latter condition does not hold for the case of the Earth: the density profile of the Earth \cite{Dziewonski1981} (Fig. \ref{Fig:Earth Density Profile}) is far from being uniform. Nevertheless, there may be situations leading to distinct spectral features, for example, if the HIO undergoes small oscillations near the center of the Earth where the density is nearly uniform. 

Sensitive gravimetry measurements are performed with a variety of instruments \cite{Hinderer2015}; among the most sensitive ones is a global
network of gravimeters, the International Geodynamics and Earth Tide
Service (IGETS \cite{IGETS,Voigt2016}) discussed in more detail in Sec.\,\ref{Sect: gravimeters}. If HIOs exist inside the Earth, each gravimeter in the network would see a weak periodic
signal at its characteristic frequencies, with a phase depending on the
geometry of the orbit and location of the gravimeter. The presence of such
frequency components in a Fourier analysis of the gravimeter time-sequence data
could indicate the presence of a HIO if it can be adequately differentiated from naturally occurring spectral features. This methodology is similar to that used by geophysicists to search for periodic oscillations of the solid core of the Earth, the so-called Slichter mode \cite{Rosat2003,ding2015slichter}.

There are also entirely different scenarios that can lead to signals in principle observable with gravimeters. For example, ultralight scalar dark-matter field can lead to effective variation of fundamental constants, including the mass of the baryons \cite{Geraci2016}. This would cause a sinusoidal variation of the Earth mass at a frequency equal to the oscillation frequency of the dark-matter field. This could be, for example, the background galactic dark matter nominally oscillating at the Compton frequency of the underlying boson \cite{Geraci2016}, an Earth-bound halo \cite{Banerjee2020}, or the field in a ``boson star" encountering the Earth \cite{Jackson2018} and leading to a transient (rather than a persistent) signal. Some such scenarios have been recently analyzed in Ref.\,\cite{Mcnally2019}. The data from the gravimeter network were also used in Ref.\,\cite{Shao2018} to set limits on a possible violation of Lorentz invariance. 

In this paper, we survey the scenarios that could potentially lead to observable effects of the HIO, discuss the sensitivity of a gravimeter network, and present a preliminary analysis of a historical record of the IGETS data. Finally, we assess the prospects of the future HIO searches based on these techniques.

During the preparation of this manuscript, we became aware of a similar work \cite{Hor2019}. While the basic idea and approach of Ref.\,\cite{Hor2019} are close to ours, the specifics are different and complementary.

\section{Capture/formation scenarios and their difficulties}
\label{Sec:capturescenarios}

One can imagine a few different scenarios in which an oscillatory gravitational signal can be produced by hidden sector objects gravitationally bound to the Earth. The most studied in the literature (see, for example, \cite{Banerjee2020,Lawson2019}) is the formation of diffuse halos of dark matter particles around the Earth.\footnote{One model suggests that axion quark nuggets (AQN) \cite{Zhitnitsky2003} explain the similarity of the dark and visible cosmological matter densities: in this model annihilation of anti-AQNs with visible matter produces a terrestrial halo of axion dark matter when AQNs hit the Earth. Although only a small fraction ($\approx10^{-17}$) of the emitted particles stay bound, the accumulation of axions over the history of the Earth can still result in a halo (see \cite{Lawson2019} for a detailed discussions of the process), although in this scenario the halo is external, virial, and of order 0.1\,kg and thus not suitable for detection with gravimeters.} One could imagine a scenario in which a meteor impact or other violent jolt set up relative oscillations between such a halo and  Earth. However, numerical models of a halo of non-interacting particles on orbits around a point
mass have shown that any overall coherent
oscillations of the halo would damp out on orbital time scales and become not observable. This can be
seen intuitively by noting that each particle in the halo has (generally) a
different orbital frequency, thus the overall oscillations will not add
coherently.

A second scenario is massive compact hidden-sector objects on orbits within, or possibly extending slightly beyond, the surface of the Earth. 
As previously noted (and as will be quantified below), objects with such trajectories do not have the enticing property of single orbit-independent orbital frequencies, and thus are in general difficult to detect by the methods discussed in this paper. 

The third scenario is massive compact objects on orbits confined sufficiently close to the center of the Earth where the Earth's density is essentially uniform. It is such objects which we will principally consider. We thus define a Hidden Internal Object (HIO) as a compact object that orbits as a single object, entirely within the Earth's core.

How could HIOs be formed? In general
this is a difficult problem for non/minimally interacting objects.  An object starting far from the Earth following a trajectory that will bring it within the interior of the Earth starts with a gravitational potential energy equal to the necessary energy to reach the Earth's escape velocity. Moreover, the velocity of generic objects in the galaxy relative to the Earth is generally on the order of the
Milky Way virial velocity of 220\,km\,s$^{\text{-1}}$, whereas the escape velocity for the Earth is 11\,km\,s$^{\text{-1}}$, so some strongly inelastic process is needed for even orbital capture, and additional energy dissipation is required to localize the object to the inside of the Earth.  Objects of interest to us might be expected, in general, to have nearly dissipationless interactions with ordinary matter. However, this does not necessarily exclude dissipation due to self-interactions or interactions with other forms of dark matter in the hidden sector. For example, one possible capture scenario is that a captured object originate from a diffuse ``cloud'' in which a small part is sheared away and gets captured by the Earth in an effective ``three-body'' collision. Such scenarios are not uncommon in celestial dynamics, where gravitational tidal forces can rip apart bound objects and capture material \cite{Roche1850}. The self-interaction scenario, however, has two serious problems.
One is that it requires the hidden-sector object to be only weakly self-bound so that some of the material can be gravitationally sheared off, but this type of weak self interaction will generically lead to either the formation of rings of the material (making our detection scheme unworkable) or to complete virialization (with the same effect). Additionally this scenario requires the matter being captured to have non-trivial self-interactions, which are generically constrained by galaxy cluster mergers \cite{Wittman2018}.

A further difficulty is that even if a dense object is captured into an Earth orbit, only extremely eccentric orbits (those intersecting the Earth) plus a dissipative interaction between the compact object and the Earth offers a plausible method for confining the object to the interior of Earth. But, such an interaction would damp the orbit until the object is confined to be mostly stationary at the center of the Earth. A possible alternative energy-loss mechanism related to gravitational polarization of the Earth is discussed in the Appendix. 

We also note that regardless of capture mechanism, any damping mechanism which is enhanced for large velocities and supressed for small velocities will tend to produce circular orbits.

We emphasize that a specific consistent scenario for HIO formation is yet to be worked out.

To give a sense of scale to our discussions of HIO masses, we
consider several quantities.
The total amount of dark matter enclosed
in a sphere with a
radius equal to that of the solar system (under the assumptions of the
standard halo model and assuming uniform density of the dark matter) is
$\approx$ \SI{3e17}{kg}, while that contained in a sphere with the
radius of the Earth is on the order of \SI{1}{kg}. Another interesting
mass to compare is obtained by considering the volume $\mathcal{V}$
traced out through the galaxy by the Earth as it has travelled through
space since its formation:
$\mathcal{V} = vAT\approx 10^{36}~{\textrm m}^3$, where $v \approx
2\times 10^5$\,m/s is the speed of the Earth relative to the galactic
rest frame, $A \approx 10^{14} \textrm{m}^2$ is the Earth's
cross-sectional area, and $T \approx 4.5 \times 10^9$ years is the age
of the Earth. Multiplying $\mathcal{V}$ by the average dark matter
density $\rho_{dm} \approx 0.4~\textrm{GeV}/c^2/\textrm{cm}^3 \approx
7\times 10^{-22}~{\textrm{kg/m}}^3$ (here $c$ is the speed of light), we
get that the total dark matter mass the Earth has passed through is about:
$\mathcal{M} \sim 10^{15}$\,kg. The feeble interactions between dark
matter and baryonic matter makes this quantity of dark matter
unachievable in normal capture scenarios 
but it can serve as an upper limit on dark matter capture. 

\section{Detection signatures}


As mentioned above, the non-uniformity of the Earth's density leads to broadening in the spectrum of the HIO orbital frequencies, nominally removing the attractive feature of the original idea that one may just look for orbits at a single unique and predictable frequency.

\begin{figure} 
\centering
\includegraphics[width=0.8\textwidth]{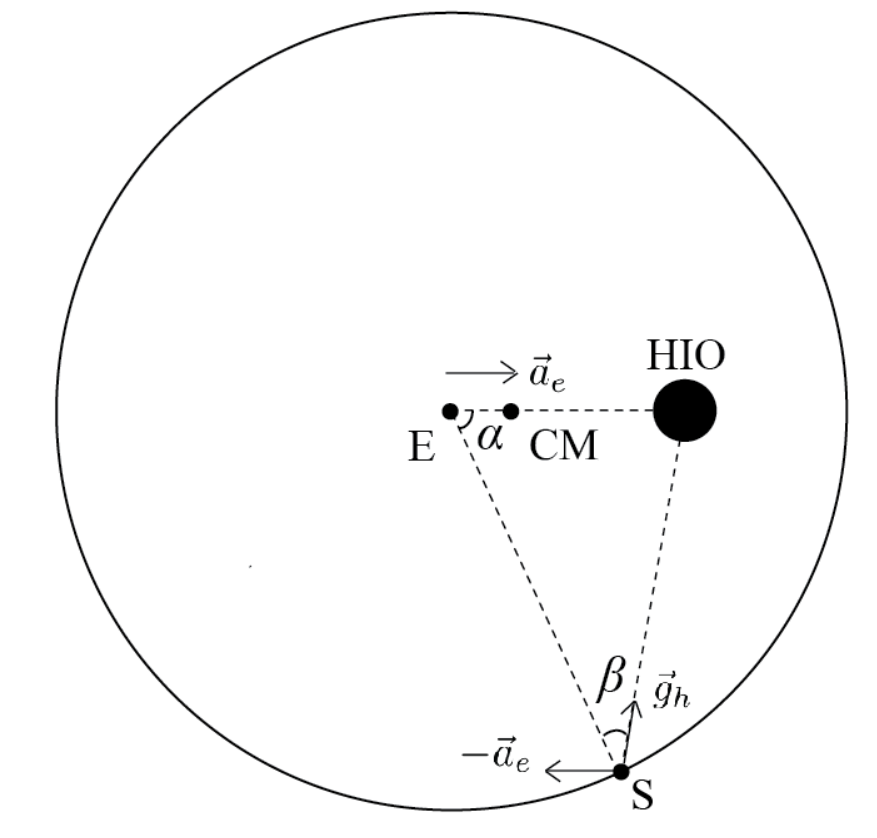}
\caption{A diagram of the contribution of the HIO to gravitational acceleration on the Earth's surface with the distance of the HIO to the center exaggerated for clarity. CM stands for the center of mass of the system, which defines the stationary frame, E is the center of the Earth and S is the location of a gravimeter station. $\vec{a}_{e}$ is the acceleration of the Earth, and $\vec{g}_{h}$ is the gravity provided by the HIO. They both contribute to the overall gravitational acceleration.}
\label{fig:Gravity_HIO}
\end{figure}

In the following derivation we will, for simplicity, assume a one-dimensional oscillation rather than the more general elliptical case. In general, we would expect an elliptical orbit (with more circular orbits being favored, as argued above), and the orbital geometry will have a non-negligible impact on the derived spectrum. However, for the sake of simplicity, we take the 1-d case as illustrative.
For small one-dimensional oscillations of a HIO near the center of the Earth the oscillation period is
\begin{equation}
   T=\frac{2\pi}{\omega_{h}}=\frac{2\pi}{\sqrt{\frac{4\pi}{3}G\rho_0}}\approx \SI{55}{min},  
\end{equation}
where $\omega_{h}$ is the angular frequency of the oscillation. In such a model, the HIO contributes to the gravitational acceleration on the Earth's surface via two terms: the gravity of the HIO ($\vec{g}_{h}$) and the acceleration of the Earth due to the HIO $\vec{a}_{e}$; see Fig. \ref{fig:Gravity_HIO}. Since the components transverse to the main acceleration of the Earth's gravity would only have second-order corrections to the readings of a scalar gravimeter, the overall effect of a HIO is reduced to:
\begin{equation}
\delta g \approx g_{h}\cos\beta + a_{e}\cos\alpha.
\end{equation}
Note that for small oscillations $\cos\beta \approx 1$, which we take to hold from here on.

Let us introduce an Earth-centered coordinate system with its $z$ axis pointing to the North Pole and denote the direction of an oscillating HIO as $(\theta_{h},\phi_{h})$. Suppose a gravimeter is placed at $(\theta_{m},\phi_{m})$ on the surface. The relationship between $\alpha$ and the Earth-centered coordinate system is: 
\begin{equation}
    \cos\alpha=\sin\theta_{h}\sin\theta_{m}\cos(\omega_{0}t+\phi_{m}-\phi_{h})+\cos\theta_{h}\cos\theta_{m},
\end{equation}
where $\omega_{0}$ is the the angular frequency of the rotation of the Earth.
Thus, the gravitational acceleration due to the HIO would contribute to the gravimeter signal as:
\begin{align}
\label{Eq:delta_g}
\delta g = &\frac{Gm_{h}}{r_{e}^{2}} + \\
&(2+\frac{\rho_{0}}{\bar{\rho}})\frac{Gm_{h}}{r_{e}^{3}}A_{h}[\sin\theta_{h}\sin\theta_{m}\cos(\omega_{0}t+\phi_{m}-\phi_{h})+\cos\theta_{h}\cos\theta_{m}]\cos\omega_{h}t,\nonumber
\end{align}
where $m_{h}$ and $A_{h}$ are the mass and amplitude of the HIO oscillation, $\bar{\rho} =$\,5.5\,g\,cm$^{\text{-3}}$ is the average density of the Earth, and $\rho_{0}=$\,13.1\,g\,cm$^{\text{-3}}$ is the density of the Earth's core. In this scenario, the HIO acts as a harmonic oscillator with a specific frequency, and the frequency is split because of the rotation of the Earth. Note however that while this splitting is exactly the sidereal frequency of the Earth's rotation in the simplified case of a 1-d non-co-rotating oscillation considered here, in general it would depend strongly on the orbital geometry (and for certain orbits be zero, for example a circular equatorial orbit).  Due to the non-uniformity of the Earth density (see Fig.\,\ref{Fig:Earth Density Profile}), this spectral pattern holds for oscillations not exceeding $\approx$\,0.1\,$r_e$ in amplitude. Note that the second term in Eq.\,\eqref{Eq:delta_g} is smaller than the first term by roughly an order of $\gamma=(2+\frac{\rho_{0}}{\bar{\rho}})\frac{A_{h}}{r_{e}}$, so assuming that $A_{h}\approx\,0.1\,r_e$, $\gamma\approx0.4$.

\begin{figure}
\label{fig:theoretical_line}
    \centering
    \includegraphics[width=\textwidth]{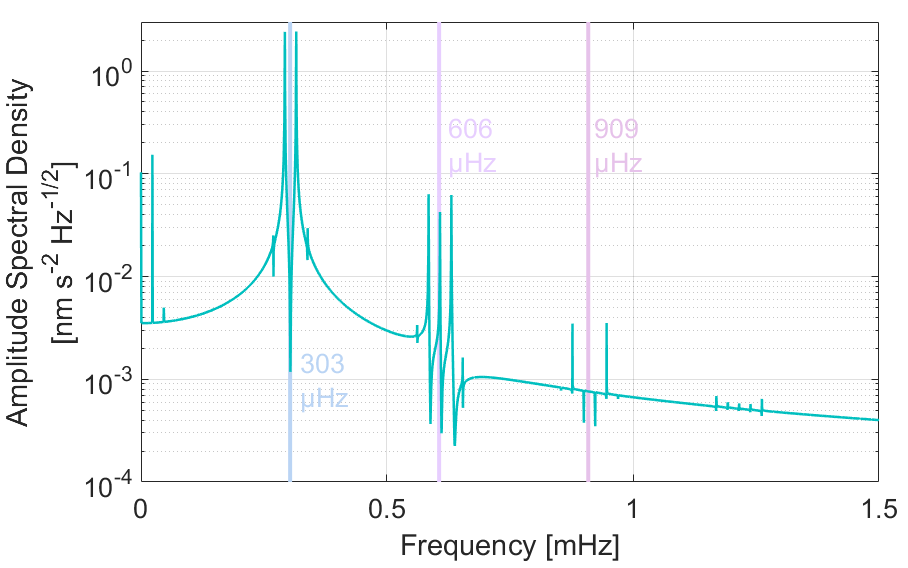}
    \caption{The theoretical spectrum of a compact HIO undergoing 1-d linear motion near the center of the Earth, on the equatorial plane. The assumed mass of the HIO here is $10^{13}\,\rm{kg}$, and the amplitude of the oscillation is 0.1\,$r_{e}$. The amplitude spectral density scales approximately as a product of the HIO mass and the the oscillation amplitude. In the spectrum, there are signals centered near the first (around 303\,$\mu$Hz) and higher (around 606\,$\mu$Hz and 909\,$\mu$Hz) harmonics due to the non-linearity of the force. Rotation of the Earth (seen as a small lowest-frequency peak) also leads to splitting of the first- and second-harmonic lines. Note that the existence and degree of the splitting is not generic, and can vary depending on the particular orbit. The calculation assumes rigid Earth but goes beyond Eq.\,\eqref{Eq:delta_g} by including transverse acceleration and the nonlinearity of the HIO motion. The nonlinearity leads to the second and higher harmonics of the spectrum, while the transverse acceleration additionally leads to the broadening of the spectral lines.}
\end{figure}

If the amplitude of such oscillation is large, there will be an amplitude-dependent frequency shift and there will appear spectral harmonics of the signal at the third and higher odd harmonics (see Fig. \ref{fig:theoretical_line}). If the orbit is elliptical and has a radius much larger than $0.1 r_e$, the motion is generally aperiodic.

In the case of the amplitude of the oscillation being less than $0.1 r_e$, any orbit is a linear combination of three orthogonal normal modes. A circular equatorial orbit in the direction of the Earth's rotation, a circular equatorial orbit opposite the Earth's rotation, and a 1-d "orbit" from the north to the south geographic poles. The 1-d polar orbit will have no frequency shift due to the Earth's rotation (as seen by an stationary observer on the Earth's surface), but the other two will have observed frequencies $\omega_{h} \pm \omega_{0}$. Therefore, a general spectrum should be expected to have three peaks at each harmonic, with the spectral triplet centered on $\omega_{h}$ and split by the Earth's rotational frequency. The relative weights of the three peaks will depend on the orbit and indeed if resolvable should uniquely determine it.

Throughout this work, we assume the Earth to be rigid, an approximation that will need to be abandoned in future, more refined analyses. If we assume that the perturbations at the relevant frequency of 0.3\,mHz propagate in the Earth with a speed of 6\,km/s (a reasonable stand-in for a typical propagation speed of an elastic wave in the Earth), then the time delay is $\approx$15\,min for the signal from a HIO to reach the gravimeter that is most sensitive to it. This is smaller than the 55\,min HIO oscillation period, but the ratio of the two times is only marginally small.
In a more refined treatment, one needs to analyze the resonant response of the Earth to determine both the amplitude and phase of the signal. If the signal frequency happens to fall near a resonant deformation mode of the Earth, this may lead to a significant modification of the phase and amplitude of the signal.
These aspects can be analysed using the 
Preliminary Reference Earth Model \cite{Dziewonski1981} in terms of the so-called Love numbers.

\section{Global network of superconducting gravimeters}
\label{Sect: gravimeters}
Superconducting gravimeters (SGs) continuously measure temporal gravity variations with high precision and long-term stability. The observations reflect the integral effects of all periodic and broadband mass variations and deformations induced by a large number of geophysical effects on temporal scales from 1\,s to several years. The long-term operation of the SGs means that all different versions by the producer GWR Instruments Inc. are currently in use, i.e., from the early commercial TT instruments to the latest transportable iGrav SGs \cite{Hinderer2015}. While the characteristics vary between the SG versions, the ultimate precision is specified by GWR for the observatory SGs as 10$^{-3}$\,nm/s$^2$ resolution in frequency domain with a noise level of better than 1\,(nm/s$^2)^2$/Hz in the seismic band from 1 to\,8 mHz and a long-term stable drift of some nm/s$^2$/a.

SG data sets are available from the database of the International Geodynamics and Earth Tides Service (IGETS) hosted by the Information System and Data Center at GFZ \cite{Voigt2016}. Originating as Global Geodynamics Project from 1997 to 2015 \cite{Crossley1999}, since 2015 IGETS provides freely accessible data from an increasing number of stations and sensors all over the world (currently 42 and 60, respectively) to support global geodetic and geophysical studies. The IGETS database provides three levels of data sets from raw gravity and local atmospheric pressure observations sampled at 1 or 2\,s (level 1) to data sets corrected for instrumental perturbations (level 2) to gravity residuals after particular geophysical corrections (level 3). Level 3 products are available for 26 stations and 36 sensors processed by EOST Strasbourg at 1\,min sampling \cite{Boy2019}. These originate from specially processed level 2 products at EOST and are reduced by gravity effects from solid Earth and ocean tides, atmospheric loading, polar motion and length-of-day variations as well as instrumental drift.

The specified SG precision of 10$^{-3}$\,nm/s$^2$ offers great possibilities in combination with stacking methods using gravity data sets from multiple stations of the IGETS network \cite{Rosat2003}. Theoretically, the sensitivity for the detection of small periodic signals could be enhanced by $\sqrt{lmn}$ assuming coherent signals with the precision increase $l$ for monthly averages as well as the total number of stations $m$ and months $n$. However, in reality, the instrumental noise from the SGs is not only superseded by station noise \cite{Rosat2018} but, above all, by a complex and significant uncertainty budget at the nm/s$^2$ level from the modelling of mass re-distributions in the atmosphere, the oceans, and hydrology on a wide range of time scales \cite{Mikolaj2019}. In addition, all SG data sets are affected by free oscillations of the Earth excited by large earthquakes in the target frequency range of 0.3\,mHz in this study \cite{Hinderer2015}.

The gravimeter precision discussed above, assuming that the geophysical effects may eventually be fully subtracted and ignoring other sources of noise, provides grounds for optimistic estimates of what one might hope to ultimately achieve in a search for HIO. For example, with a network with a similar number of stations as the existing one and assuming signals from the stations are added coherently and with on the order of a month of averaging time, the cumulative sensitivity could, in principle, reach on the order of $10^{-7}$\,nm/s$^{\text{2}}$. 
Taking an average oscillation amplitude of $A_{h}=$\,0.1\,$r_e$\,($\approx$\,637\,km), the smallest detectable mass of the network would be:

\begin{equation}
\label{eq.5}
    m_{min}=(2+\frac{\rho_{0}}{\bar{\rho}})^{-1}\frac{\delta g r_{e}^3}{GA_{h}} = \SI{1e8}{kg}.
\end{equation}
This mass can be compared to the reference values discussed at the end of Sec.\,\ref{Sec:capturescenarios}.

\section{Analysis and preliminary results}
\label{Sec:Analysis and results}
\begin{figure}
    \centering
    \includegraphics[width=\textwidth]{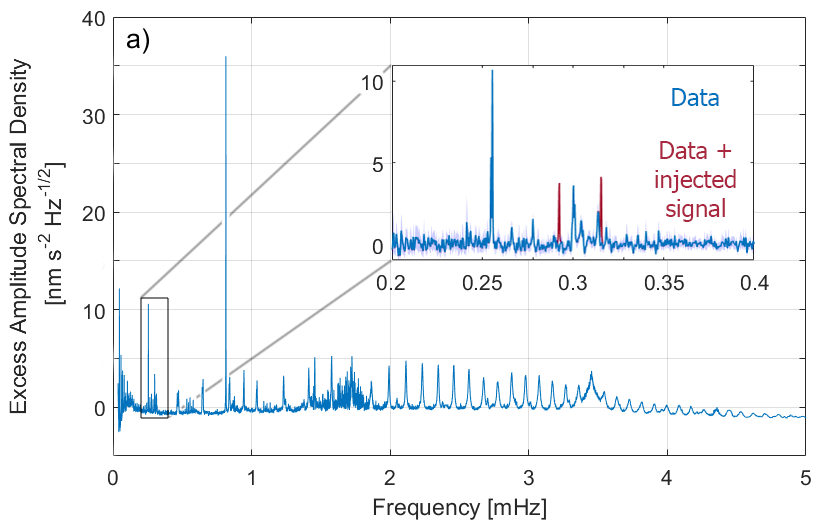}    \caption{(a) The amplitude spectral density of the IGETS level 3 data sets, with baseline removal performed after the averaging. The large spike around 800\,$\mu$Hz is due to the $_0 S _0$ ``breathing" mode of Earth \cite{Rosat2007,Dziewonski1981,Masters1995}.
    The inset (b) shows details around 303\,$\mu$Hz, where the signal from a HIO orbiting near the center of the Earth would lie. The dark red line corresponds to the data with the minimum-detectable signal ($m=10^{14}$\,kg) injected into it.}

    \label{fig:fft_inset}
\end{figure}

Our analysis technique involves taking the periodogram of each one-month block of data from all stations and averaging to obtain an estimated global power spectral density (Bartlett's method). We then compute the Amplitude Spectral Density (ASD) from this and  remove the baseline by fitting to a function of the form:
\begin{equation}
    y(f) = \frac{A}{f} + \frac{B}{f^2}+ \frac{C}{f^3} + \frac{D}{f^4} + y_0\,.
\end{equation}
We only fit in the region of higher frequency than tidal effects ($> 36.55$ $\mu$Hz), to avoid tide-induced fit artifacts. This results in a single averaged spectrum; 
see Fig.\,\ref{fig:fft_inset} a). 
Some of the spectral features seen in the spectrum are well-known in geophysics. For example, the sharp spike around 800\,$\mu$Hz is the fundamental radial mode $_0$S$_0$ \cite{Rosat2007}. Other modes in the region of interest have been reported (see, for example, Fig.\,21 and the corresponding discussion of the mode identification in Ref.\,\cite{Hinderer2015}).
There are also features in the data that do not generally appear in the raw gravity data from superconducting gravimeters and are possibly artifacts from reduction models used to obtain the level 3 data. This will be subject of further investigation.

We looked for a HIO by fitting three Lorentzians, one centered around 303\,$\mu$Hz, and the others centered around 303\,$\mu$Hz $\pm$ 11.6 $\mu$Hz (corresponding to a period of  1 day) to the data, and did not detect a signature above noise consistent with the HIO scenario we have presented. To estimate the minimum detectable HIO mass oscillating with an amplitude of $0.1 r_e$, we injected a signal (from our simulations presented in Fig. \ref{fig:theoretical_line}) into the data. With this, we determined that the minimum injected signal that is detected corresponds to a HIO mass of $\approx10^{14}$\,kg at this radius. Details of the analysis can be found in the Appendix. Note that this is not a traditional exclusion limit of a mass, but of a mass on a specific orbit.


This upper bound on the signal, assuming an oscillation $\SI{637}{km}$ ($\approx$ 0.1$r_e$), would  corresponds to an acceleration of $\delta g$ of
\begin{equation}
\delta g = (2+\frac{\rho_{0}}{\bar{\rho}})\frac{mGA_h}{r_e^3} = \SI{72}{pm/s^{2}}.
\end{equation}

Although this upper bound on the mass oscillating with an amplitude of $0.1r_e$ is relatively small ($\approx$\,$10^{-10}$ of Earth's mass), the current sensitivity is still several orders of magnitude short of the estimated sensitivity in Eq.\,\eqref{eq.5}. This difference arises from various factors. Although the contributions of the instrument drift and pressure are largely removed already, the employed data-fix techniques may introduce errors. Also, there are tides and nontidal loading factors, rainfall and other hydrological factors, station disturbances, seismic factors, and other natural and anthropological contributions to the overall acceleration. Tides generally produce clear, discrete lines in the spectra that are harmonics of the tidal frequency (period $\approx 12\,$hrs), which have little impact on the region of interest (period $\approx 55\,$min). As for the seismic influences, there are both persistent oscillations, which are a response to other periodic driving forces, which can be roughly evaluated by the Earth model and transient incidences (for example, earthquakes), which, along with other transient factors, can be removed by data selection. Data selection is also an effective technique for dealing with nontidal and hydrological loading factors which can have an effect of as big as $\approx \SI{1e4}{nm/s^{2}}$ over a few days. These effects produce noise-like spectra, deteriorating the sensitivity of the network to HIO signals \cite{Hinderer2015}.

In the future stages of this work, one should be able to enhance the precision substantially by
performing a phase sensitive analysis. Furthermore, the phase information would
enable us to approximately determine the specific orbit of a HIO within the
Earth, as the phases of a gravimeter signal would depend on its location on Earth's surface. 


The stated HIO-mass sensitivity of $10^{8}$\,kg  is based on the assumption that the signals of different stations are completely correlated, and neglects all noise sources other than those of the sensors. 
In our case where HIO move as a single object inside the Earth, the signal obtained by different stations are indeed correlated. However, if other possible scenarios are to be considered, then the correlation would be incomplete and the sensitivity would deteriorate.

We note that correlation-analysis techniques are currently employed by the existing sensor networks such as LIGO/VIRGO \cite{LIGO2019} for the detection of gravitational waves, as well as by magnetometer (GNOME \cite{Afach2018}) and clock (for example, GPS.DM \cite{Derevianko2018}) networks for the detection of the galactic dark matter. 





\section{Conclusions and outlook}

We have analyzed various possible scenarios of hidden gravitationally bound objects that have weak or no interactions (other than gravitation) with normal matter, and have discussed their possible influence on the total gravitational field measured at the surface of the Earth. With data sets from IGETS, we used Fourier analysis to search for characteristic spectral lines that could be an indication of the existence of such objects. Although no evidence has been found, we estimate that the smallest detectable mass using the current network could perhaps ultimately reach as low as $\approx\SI{1e8}{kg}$. Such hidden gravitationally bound objects could potentially be related to nonbaryonic dark matter.

An alternative scenario to HIO trapped in the Earth is a change of the Earth's gravitational field under the influence of some background bosonic field, for example, that due to dilatons.
Serving as dark matter candidates, dilatons and other bosonic fields can have linear interactions with nucleons, changing their effective masses at the Compton frequency associated with the mass. Under appropriate conditions, the mass of the Earth could oscillate slightly at the particle Compton frequency \cite{Geraci2016}. The superconducting gravimeter data considered in this note are resampled to the once per minute cadence, so the
smallest frequency that can be detected is \SI{1/120}{Hz}, corresponding to a particle
mass of \SI{3.5e-17}{eV}. The modification of the Earth mass by the presence of an oscillating field  $\phi=\phi_{0}\cos\omega_{\phi} t$ is described by 
\begin{equation}
    m_{eff}=(1+\frac{\sqrt{\hbar c}\phi}{\Lambda_{1}})m_{e},
\end{equation}
where $m_{e}$ is the mass of the Earth, $\omega_{\phi}$ is the frequency of the oscillating field, and $\phi_{0}=\hbar\sqrt{2\rho_{DM}}/(m_{\phi}c)$ is related to the mass of the bosonic particle $m_{\phi}$ and the local density of dark matter $\rho_{DM}$. $\Lambda_1$ is the coupling constant averaged over all the Earth's atoms.

Assuming the optimistic sensitivity of $10^{-17}\,g\approx 10^{-7}$\,nm/s$^2$ discussed in Sec.\,\ref{Sect: gravimeters}, the network can detect such variance if
\begin{equation}
    m_{\phi}c^{2}\Lambda_{1}\leq \SI{2.5e14}{eV^{2}},
\end{equation}
which is compatible to the sensitivity of future atom interferometers discussed in Ref.\,\cite{Geraci2016} and performing better than current equivalence-principle tests if $m_{\phi}\leq\SI{1e-18}{eV}$. Additionally, periodic mass variations of Earth's mass could appear as sidebands in Earth's vibrational modes.


Another possibility is that there could be ``boson stars'' encountering the Earth that affect the gravitation field. Supposing such influence is only detectable when the distance between the star and the Earth is closer than $10r_{e}$ and taking the characteristic relative velocity of the Earth and the boson star as the galactic virial velocity $\approx 10^{-3}c$, such transient signal would possibly last $\approx$\SI{5}{min}, which means that its timing is ideal for detection using IGETS's level-2 and level-3 data sets. According to an estimate in \cite{Jackson2018}, the maximum acceleration felt during an encounter is $10^{-19}\,g$, so a significant improvement in sensitivity would be needed if one is to detect such events using gravimeters.

In conclusion, advanced gravimeter networks could be useful for detection of exotic matter and  future improvements in the hardware and, particularly, in advanced data analysis may enable mounting competitive searches.


\section*{Acknowledgements}
The authors acknowledge helpful discussions with Joshua Eby, Ernst M. Rasel, Surjeet Rajendran, and the members of the CASPEr and GNOME collaborations. This work was supported in part by the European Research Council (ERC) under the European Union Horizon 2020 research and innovation program (grant agreement No 695405) and by the DFG via the Reinhart Koselleck project and DFG Project ID 390831469:  EXC 2118 (PRISMA+ Cluster of Excellence) and by the U.S. National Science Foundation under Grants No. PHY-1707875 and PHY-1505903.

\section*{Appendix A: tidal ring-down}
\label{Appendix:tidal_capture}
When an object passes through the Earth, its gravity causes the elastic energy of the Earth to change. 
Part of this energy is dissipated through friction,
which could in principle be an indirect method contributing to energy loss of a hidden object.

Consider a point object with mass $M$ crossing the Earth with an initial velocity of $v_{0}$.
When the object is at the center of the Earth, the elastic energy of the Earth is maximized. The pressure increment $p(r)$ due to the object inside the Earth can be found from the force balance:
	\begin{equation}
	\label{eq.force balance}
p(r)4\pi r^{2}-p(r+dr)4\pi (r+dr)^{2}=\frac{4GM\rho\pi r^{2}dr}{r^{2}}~.
	\end{equation}
Here, the density of the Earth $\rho$, for simplicity of the argument, is assumed to be uniform.
	Rewriting Eq.\,\eqref{eq.force balance} as a differential equation with boundary conditions:
	\begin{equation}
	-r^{2}\frac{dp}{dr}-2pr=GM\rho,~~p(r_{e})=0~,
	\end{equation}
we find the solution:
	\begin{equation}
	p=\frac{GM\rho(r_{e}-r)}{r^{2}}~.
	\end{equation}
The deformation of the Earth and the pressure increment can be related via the Young Modulus $E$:
	\begin{equation}
	E\frac{\delta r}{dr}=p,
	\end{equation}
where $\delta r$ is the compression of the layer $dr$.
The total elastic energy:
	\begin{equation}
	W=
\int_{0}^{r_{e}}\frac{1}{2}p4\pi r^{2}\delta r=\frac{2\pi(GM\rho)^{2}}{E}\int_{0}^{r_{e}}\frac{(r_{e}-r)^{2}}{r^{2}}dr.
	\end{equation}
This energy diverges at the lower limit of the integral meaning that we must impose some physical cut-off. We can, for instance, assume that 	
the core of the Earth is incompressible and impose a cutoff at $r=r_{c}\approx \SI{1221.5}{km}$, where $r_{c}$ is the core radius (another possible cutoff would come from the finite size of the object $M$). For $r_{c}\ll r_{e}$, we have:
	\begin{equation}
	\label{eq.elastic energy}
	W=\frac{2\pi(GM\rho)^{2}}{E}\frac{r_{e}^2}{r_{c}}~.
	\end{equation}
The information on the friction inside the Earth can be obtained from seismic data, specifically from the measured $Q$ factors of the seismic oscillations. According to the PREM model \cite{Dziewonski1981}, a typical $Q$ factor of seismic waves is $Q \approx 10^{3}$. If $M=\SI{1e12}{kg}$, the energy loss per pass of an object through the Earth would be roughly
	\begin{equation}
	E_{f}\approx \frac{W}{Q}\approx \SI{1e7}{J}~.
	\end{equation}
For comparison, the escape energy of such mass is:
	\begin{equation}
	E_{escape}=\frac{1}{2}Mv_{escape}^{2}\approx \SI{6e19}{J}~,
	\end{equation}
so the energy loss is far too small for what is needed to capture the object. Somewhat less pessimistic numbers may be obtained, as can be seen from Eq.\,\eqref{eq.elastic energy} for larger masses $M$ and by taking a smaller cut-off radius. Note that Earth's resonances may significantly modify the considerations above.

If the mechanism were to work, an attractive feature is that as the object winds down towards the core, the energy losses decrease because for small oscillations near the Earth center, the energy losses will become quartic in the oscillation amplitude, so the losses effectively turn off, enabling long ``ringing'' of the oscillation.


\section*{Appendix B: data analysis validation}
To obtain the smallest detectable HIO signal, we inject a synthetic signal into our data set. We vary the amplitude of the injected signal, and find the smallest amplitude we can reliably detect. This amplitude is then converted to a HIO mass in a optimistic orbit (that is, one with $r=0.1r_e$, and a favorable orbital geometry). In this sense we do not place an exclusion limit on the mass of the HIO, but on a combination of mass and orbital parameters. 

The injected signal corresponds to the time-domain acceleration whose spectrum was presented in Fig. \ref{fig:theoretical_line}. The signal was scaled by the corresponding mass and injected into the raw time-domain gravity residuals of the SG data sets, with the same phase and amplitude in each injection. The appropriate spectral densities are then computed as normal, by computing and then averaging periodograms. We fit the result to three Lorentzian functions (one at 303 $\mu$Hz, the other two at 303 $\pm$ 11.6 $\mu$Hz. They all have an identical width. This width and the amplitudes are free parameters) using the Nelder-Mead Simplex minimization algorithm and a $\chi^2$ cost function. Per-point errors were estimated as the standard deviation of the mean of each frequency point across the periodograms.



\begin{figure}
    \centering
    \includegraphics[width=  1 \textwidth]{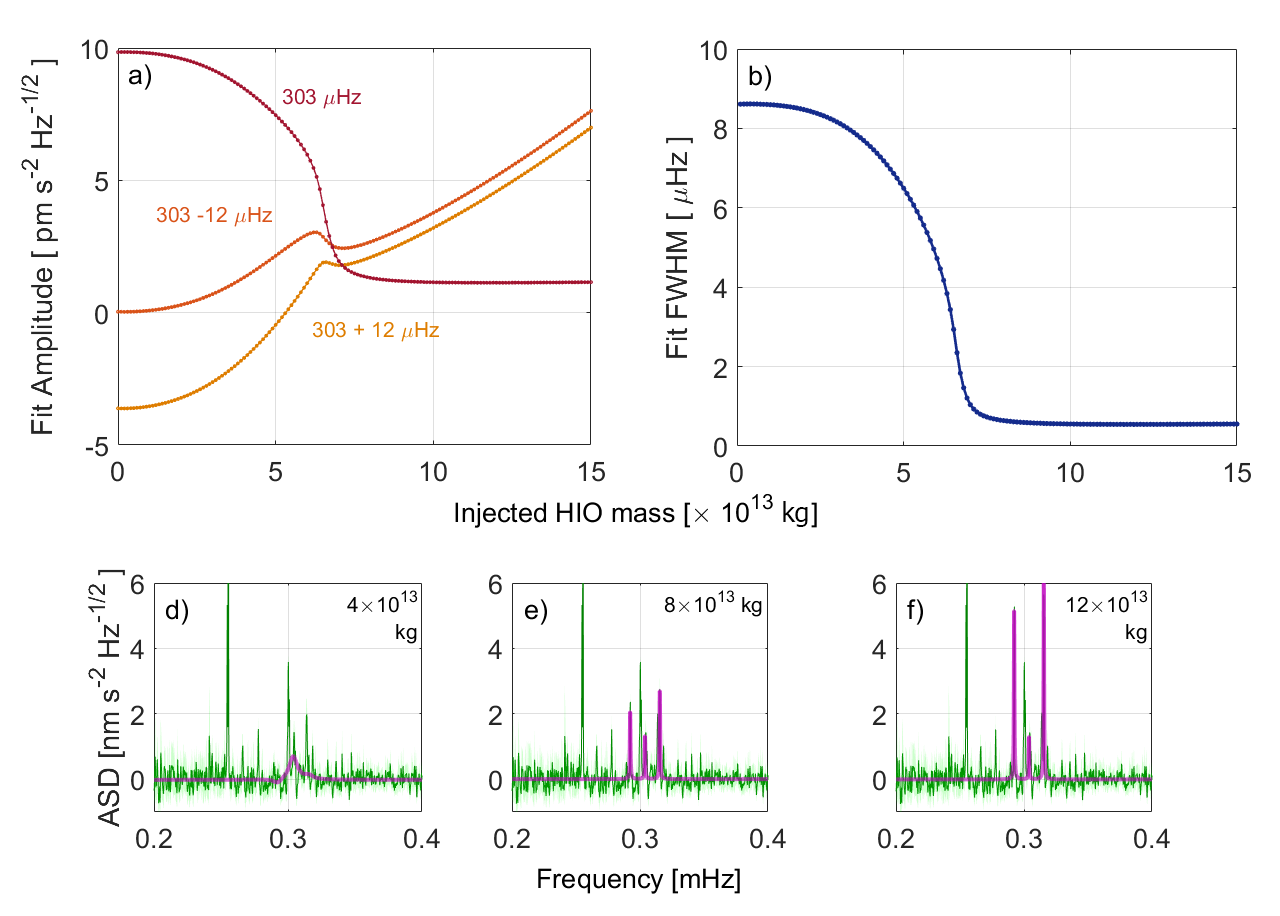}    \caption{Fitting scenarios for different injected amplitudes. (a) and (b) show the behaviour of the fit amplitudes and the width (full width at half maximum, (FWHM)) for the three Lorentzians at different injected signals. (c,d,e) are examples of the fit results for injected signals corresponding to a HIO mass of $4\times10^{13}$, $8\times10^{13}$ and $12\times10^{13}$\,kg, respectively. The green lines are the data, the light green shaded region represents the standard error of the mean and the purple line corresponds to the fit results. }

    \label{fig:appendix}
\end{figure}

The results of the fitting are in Fig. \ref{fig:appendix}, and it is apparent when the injected signal becomes detectable, at around 8$\times$10$^{13}$\,kg. At this point, the width and the  amplitude at 303\,$\mu$Hz converge to a single value, and the fit amplitudes at 303\,$\pm$\,$\mu$Hz begin to be proportional to the injected amplitude. This is expected considering the relative amplitudes of the peaks in the injected signal (see Fig. \ref{fig:theoretical_line}).

We therefore set a limit at a combination of HIO mass and orbital dynamics resulting in from a HIO mass of $10^{14}$\,kg, corresponding to a gravitational signal strength of $\approx$\SI{72}{pm/s^2}. We note that given that this sensitivity is computed using a specific injected signal, it could in principle vary depending on the injected orbit. Future analysis should average over spectra produced by a large number of simulated orbital geometries.

\newpage

\bibliographystyle{unsrt}
\bibliography{Gravimeter}

\begin{thebibliography}{10}

\bibitem{Bertone2018}
Gianfranco Bertone and Dan Hooper.
\newblock History of dark matter.
\newblock {\em Rev. Mod. Phys.}, 90:045002, Oct 2018.

\bibitem{Dziewonski1981}
Adam~M. Dziewonski and Don~L. Anderson.
\newblock Preliminary reference earth model.
\newblock {\em Physics of the Earth and Planetary Interiors}, 25(4):297 -- 356,
  1981.

\bibitem{Hinderer2015}
J.~Hinderer, D.~Crossley, and R.J. Warburton.
\newblock 3.04 - {S}uperconducting {G}ravimetry.
\newblock In Gerald Schubert, editor, {\em Treatise on Geophysics (Second
  Edition)}, pages 59 -- 115. Elsevier, Oxford, second edition, 2015.

\bibitem{IGETS}
The web page of international geodynamics and earth tide service.
\newblock \url{http://igets.u-strasbg.fr/index.php}.

\bibitem{Voigt2016}
Christian Voigt, Christoph F{\"o}rste, Hartmut Wziontek, David Crossley, Bruno
  Meurers, Vojtech P{\'a}link{\'a}{\v{s}}, Jacques Hinderer, Jean-Paul Boy,
  Jean-Pierre Barriot, and Heping Sun.
\newblock {\em Report on the data base of the international geodynamics and
  earth tide service (IGETS)}.
\newblock Deutsches {GeoForschungsZentrum GFZ}, 2016.

\bibitem{Rosat2003}
Severine Rosat, Jacques Hinderer, David Crossley, and Luis Rivera.
\newblock The search for the {S}lichter mode: comparison of noise levels of
  superconducting gravimeters and investigation of a stacking method.
\newblock {\em Physics of the Earth and Planetary Interiors},
  140(1-3):183--202, 2003.

\bibitem{ding2015slichter}
Hao Ding and Benjamin~F Chao.
\newblock The {S}lichter mode of the earth: Revisit with optimal stacking and
  autoregressive methods on full superconducting gravimeter data set.
\newblock {\em Journal of Geophysical Research: Solid Earth},
  120(10):7261--7272, 2015.

\bibitem{Geraci2016}
Andrew~A. Geraci and Andrei Derevianko.
\newblock Sensitivity of atom interferometry to ultralight scalar field dark
  matter.
\newblock {\em Phys. Rev. Lett.}, 117:261301, Dec 2016.

\bibitem{Banerjee2020}
Abhishek Banerjee, Dmitry Budker, Joshua Eby, Hyungjin Kim, and Gilad Perez.
\newblock Relaxion stars and their detection via atomic physics.
\newblock {\em Communications Physics}, 3(1):1--8, 2020.

\bibitem{Jackson2018}
D.~F. Jackson~Kimball, D.~Budker, J.~Eby, M.~Pospelov, S.~Pustelny,
  T.~Scholtes, Y.~V. Stadnik, A.~Weis, and A.~Wickenbrock.
\newblock Searching for axion stars and {Q}-balls with a terrestrial
  magnetometer network.
\newblock {\em Phys. Rev. D}, 97:043002, Feb 2018.

\bibitem{Mcnally2019}
Rees~L McNally and Tanya Zelevinsky.
\newblock Constraining domain wall dark matter with a network of
  superconducting gravimeters and ligo.
\newblock {\em arXiv preprint arXiv:1912.06703}, 2019.

\bibitem{Shao2018}
Cheng-Gang Shao, Ya-Fen Chen, Rong Sun, Lu-Shuai Cao, Min-Kang Zhou, Zhong-Kun
  Hu, Chenghui Yu, and Holger M\"uller.
\newblock Limits on {L}orentz violation in gravity from worldwide
  superconducting gravimeters.
\newblock {\em Phys. Rev. D}, 97:024019, Jan 2018.

\bibitem{Hor2019}
C.~J. Horowitz and R.~Widmer-Schnidrig.
\newblock Gravimeter search for compact dark matter objects moving in the
  earth; ar{X}iv 1912.00940, 2019.

\bibitem{Lawson2019}
Kyle Lawson, Xunyu Liang, Alexander Mead, Md~Shahriar~Rahim Siddiqui, Ludovic
  Van~Waerbeke, and Ariel Zhitnitsky.
\newblock Gravitationally trapped axions on the {E}arth.
\newblock {\em Phys. Rev. D}, 100:043531, Aug 2019.

\bibitem{Zhitnitsky2003}
Ariel~R Zhitnitsky.
\newblock `nonbaryonic’ dark matter as baryonic colour superconductor.
\newblock {\em Journal of Cosmology and Astroparticle Physics},
  2003(10):010–010, Oct 2003.

\bibitem{Roche1850}
Edouard Roche.
\newblock La figure d’une masse fluide soumise l’attraction d’un point
  loign.
\newblock {\em Acad. des sciences de Montpellier}, 1:1847--50, 1850.

\bibitem{Wittman2018}
David Wittman, Nathan Golovich, and William~A Dawson.
\newblock The mismeasure of mergers: revised limits on self-interacting dark
  matter in merging galaxy clusters.
\newblock {\em The Astrophysical Journal}, 869(2):104, 2018.

\bibitem{Crossley1999}
D~Crossley, Jacques Hinderer, G~Casula, O~Frnacis, H-T Hsu, Y~Imanishi,
  G~Jentzsch, J~K{\"a}{\"a}ri{\"a}nen, J~Merriam, B~Meurers, et~al.
\newblock Network of superconducting gravimeters benefits a number of
  disciplines.
\newblock {\em Eos, Transactions American Geophysical Union}, 80(11):121--126,
  1999.

\bibitem{Boy2019}
J.-P. Boy.
\newblock Description of the level 2 and level 3 {IGETS} data produced by
  {EOST}; https://isdc.gfz-potsdam.de/igets-data-base/documentation/, 2019.

\bibitem{Rosat2018}
S~Rosat and J~Hinderer.
\newblock Limits of detection of gravimetric signals on {E}arth.
\newblock {\em Scientific {R}eports}, 8(1):15324, 2018.

\bibitem{Mikolaj2019}
M~Mikolaj, M~Reich, and A~G{\"u}ntner.
\newblock Resolving geophysical signals by terrestrial gravimetry: A time
  domain assessment of the correction-induced uncertainty.
\newblock {\em Journal of Geophysical Research: Solid Earth},
  124(2):2153--2165, 2019.

\bibitem{Rosat2007}
Severine Rosat, Shingo Watada, and Tadahiro Sato.
\newblock Geographical variations of the $_0$s$_0$ normal mode amplitude:
  predictions and observations after the {S}umatra-{A}ndaman earthquake.
\newblock {\em Earth, planets and space}, 59(4):307--311, 2007.

\bibitem{Masters1995}
TG~Masters and R~Widmer.
\newblock Free oscillations: frequencies and attenuations.
\newblock {\em Global Earth Physics: a handbook of physical constants}, 1:104,
  1995.

\bibitem{LIGO2019}
LIGO~Scientific Collaboration, Virgo Collaboration, et~al.
\newblock A guide to {LIGO-Virgo} detector noise and extraction of transient
  gravitational-wave signals.
\newblock {\em arXiv preprint arXiv:1908.11170}, 2019.

\bibitem{Afach2018}
S.~Afach, D.~Budker, G.~DeCamp, V.~Dumont, Z.D. Gruji{\'c}, H.~Guo,
  D.F.~Jackson Kimball, T.W. Kornack, V.~Lebedev, W.~Li, H.~Masia-Roig, S.~Nix,
  M.~Padniuk, C.A. Palm, C.~Pankow, A.~Penaflor, X.~Peng, S.~Pustelny,
  T.~Scholtes, J.A. Smiga, J.E. Stalnaker, A.~Weis, A.~Wickenbrock, and
  D.~Wurm.
\newblock Characterization of the global network of optical magnetometers to
  search for exotic physics ({GNOME}).
\newblock {\em Physics of the Dark Universe}, 22:162 -- 180, 2018.

\bibitem{Derevianko2018}
Andrei Derevianko.
\newblock Detecting dark-matter waves with a network of precision-measurement
  tools.
\newblock {\em Phys. Rev. A}, 97:042506, Apr 2018.

\end{thebibliography}

\end{document}